# Patience and Impatience of Traders: Formation of the Value-At-Price Distribution Through Competition for Liquidity

Peter Lerner[1]

Abstract

An ability to postpone one's execution without much penalty provides an important strategic advantage in high-frequency trading. To quantify competition between traders one has to resort to a consistent theory of formation of the execution price from market expectations and quotes. This theory was provided in 2005 by Foucault, Kadan and Kandel. I derive asymptotic distribution of the bids/offers as a function of the ratio of patient and impatient traders using my modification of the Foucault, Kadan and Kandel dynamic Limit Order Book (LOB) model.

My modification of the LOB model allows stylized but sufficiently realistic representation of the trading markets. In particular, dynamic LOB allows simulation of the distribution of execution times and spreads from high-frequency quotes. Significant analytic progress is made towards understanding of trading as competition for immediacy of execution between traders. The results are qualitatively compared with empirical volume-at-price distribution of highly liquid stocks.

1) Scitech Analytical Services, LLC Woodland Drive, State College, PA 16803, Penn State U. (affiliate), pblerner@syr.edu. I am thankful to the members of the finance seminar at Illinois Institute of Technology for their valuable comments on the first (2012) version and to my discussants at Global Finance Conference (Monterey, May 2013), Hisham Farag (U. Birmingham) and the 5th Meeting on the Behavioral Finance and Economics (Chicago, Sept. 2013), Mikaela Pagel (Berkeley). My thanks are extended to Lorne Switzer (Concordia U.), chair at the Monterey Meeting, for his advice during and after the conference, especially for using the WECO Bloomberg(C) function for economic events. I bear responsibility for all errors.

**Introduction**

The propagation of high-frequency trading (HFT), along with the algorithmic executions and extreme events usually associated with HFT such as the "flash crash" of May 6, 2010 and several interruptions of Nasdaq trading in 2013, have brought financial microstructure studies from the backburner to a forefront of economic theory.[2] However, the theoretical reasons for abnormal reactions of electronic trading systems on quickly changing supply and demand conditions, as well as beliefs and preferences, remain mysterious. <u>Thus, it is important to have dynamic equations, however simplified, to be able to model transitional regimes in the formation of bid-ask prices within trading systems.</u>

Currently, a large number of microstructure models, each with their own level of realism, empirical accuracy and analytic tractability, are in existence. In general, microstructure models are difficult subjects for empirical verification because they frequently express unobservable parameters (such as asset volatility or microstructure noise) in terms of other unobservable parameters such as the fraction of informed traders (O'Hara, 1995, Hasbrouck, 2007, Lerner 2009). Yet, they are necessary if we are ever to move beyond random walk prices and perfectly efficient markets. The report of the Swiss Finance Institute proclaims: "One of the most pressing subjects is to come up with a realistic agent-based model, where crisis and complexity arise from simple rules and interactions in a universal way." (Sornette, von der Becke, 2011)

The simplest "behavioral" effect in market microstructure is the distinction between patient and impatient traders. It is well known that, in the absence of changing economic fundamentals, the traders who have freedom to postpone their execution can receive a more favorable execution price than those who have no such opportunity. (Harris, 2003) Yet, to quantify this effect of micro-liquidity one has to resort to a consistent theory of formation of the execution price from market expectations and quotes.

---

[2] For instance, the preprint of "Flash Crash: Flow Toxicity, Liquidity Crashes and the Probability of Informed Trading" was downloaded from ssrn.com more than 15,000 times by January 21, 2013 from the date of its release in 2011. (Easley, Lopez Prado and O'Hara, 2011)

In the current paper we propose a theory based on the general assumption that trading can be represented as a random walk of successive bids (offers) over the state space of LOB. This insight allows me to replace a very complicated process price with a family of 1-D random walks. This approach was pioneered by Wang and Obizhaeva, (2005) in the context of optimal execution of quotes. Our approach is completely independent of Wang and Obizhaeva but instead makes a dynamic extension of the static model provided in the same year 2005 by Foucault, Kadan and Kandel (FKK).

Our paper is structured as follows. The first section is a literature review. Second and third sections discuss empirical data on Volume-at-Average-Price (VWAP) distributions. The fourth section is an exposition of the FKK theory. The fifth section is our own dynamic extension of the FKK model. The sixth section compares simulated distributions with VWAP. The seventh section discusses biological and social analogies of the proposed theory. In the Section 8, we make comments concerning numerical size of the dynamical limit order book effects. The paper ends with conclusion and has three appendices dealing with analytical features of the model.

**1. Literature review**

There are several, mutually complementary approaches to market microstructure. One of them is a dynamic modeling of the Limit Order Book (LOB), which is described in Chapter 8 of the book by De Jong and Rindi (2010). The most analytically tractable of these models has been proposed by Foucault, Kadan and Kandell (FKK) in 2005 (Foucault, Kadan and Kandell, further FKK 2005). FKK can be placed in a class of barrier-diffusion models first proposed by Harris (1998). In this class of model, the order submission evolves as diffusion and the limit orders execute when the price hits a barrier (Hasbrouck, 2007).

The FKK model describes LOB in terms of two state variables: execution time $T_h$ and the proportion of so-called "patient" and "impatient" traders, $\theta_P$ and $(1-\theta_P)$. In the FKK model, all trading is in immediacy. Namely, patient traders benefit at the expense of impatient traders who cannot wait for a more favorable price for execution of their trades.

This heuristic and intuitive picture of trading is supported by the empirical results of Menkveld (2010) on the HFT. Namely, Menkveld identified the role of HFT, i.e., supposedly well-informed and well-connected traders as a substitute, electronic version of the market maker (see also Foucault, 2012). That line of research finds its further development in Jovanovic and Menkveld's (2011) study of HFT as latter-day middlemen.

Execution time is difficult to observe and the proportion of impatient traders is completely unobservable. While, recently, exchanges and "dark pools" started to provide information on all limit orders with their time stamps on a proprietary basis, this information is unlikely to be available soon for commodities, which are mostly transacted on a party-to-party basis (see e.g. Skjeltorp, 2012). Skjeltorp, Sojli and Wah Tham (2012) performed the first empirical tests of FKK. But even they acknowledged that direct clarification of liquidity externalities is "extremely" challenging. In real markets, liquidity is determined endogenously—while in most models it can be introduced as a functional of the state variables—and there are few good instruments with which one can identify this functional (see also Barclay and Henderschott, 2004 and Henderschott and Jones, 2005).

Skjeltorp, Sojli and Wah Tham (2012, further SST) built a complete LOB from Nasdaq OMX BX data with millisecond accuracy. Thus, in view of the results of SST, the necessity for a simplified dynamic model as in this paper might seem obviated, but there are extenuating circumstances.

First, the increased proliferation of the "dark pools" as alternative venues for execution (see, e.g. Angel, Harris and Spatt, 2010) makes it unlikely that LOBs of most markets will become directly accessible. Furthermore, recent studies (Switzer and Fan, 2010) demonstrate that the most frequently used measures of trading costs such as bid-ask spreads and volumes (but not the number of trades) have very low instantaneous correlation with actual cost of trading. In the view of the Switzer and Fan study, it seems advantageous to develop an endogenous theory of formation of the bid-ask spread independently of the empirical metrics involved.

Second, one cannot exclude the emergence of derivative instruments contingent on liquidity. Finally, a relatively parsimonious analytic model can be important despite oversimplifications.

An alternative possibility for the empirical verification of a dynamic LOB model would be to develop functionals of these state variables, which simulate the behavior of the observable quantities related to prices and volumes of traded securities. Once the probability distribution (measure) of the price-by-volume has been established, its integrals provide expectations for market-observed values (see Section 4). This approach has an advantage over the time-series analysis (Aït-Sahalia, 2012, Aït-Sahalia, 2013) in that the integrated quantities are robust with respect to the jumps of the state variables and particular statistics of the microstructure noise.

In this paper, I derive a dynamic distribution for the evolution of the proportion of impatient traders in time. But before we can discuss the observable consequences of the FKK, we must provide a brief exposition of this theory, for which we shall provide a dynamical version. In exposition of the FKK, we mostly follow De Jong and Rindi (2009).

## 2. Empirical evidence on integrated volume-at-price distributions

My theory is proposed to describe one stylized fact, namely, that the VWAP (Volume-at-Weighted-Average-Price) distributions of liquid stocks integrated over several ordinary days have rather similar shapes. Specifically, during a single day, they typically form a unimodal structure, which crudely resembles a normal or lognormal distribution (Fig. 1).

[place Figure 1 approximately here]

During the next few days, a peak in the distribution widens and forms into a random structureless shape (Fig. 2a). Finally, after more than one week, a typical VWAP settles into a bimodal distribution (Fig. 2b).

[place Figure 2 approximately here]

In Fig. 3, I show Volume-at-Weighted-Average-Price (VWAP) distributions of quotes for Microsoft stock at Nasdaq during 1, 5 and 10 days of information collection in mid-August 2011, respectively. Ten days of collecting was the practical limit of the Bloomberg© VWAP function for that time and a longer observation interval was impossible. During these days in August, the trading volume was between 50 and 100 million shares. Cumulative data on trading are collected in Table 1.

Table 1. **Descriptive statistics of trading in Microsoft stock during the period 08/11/11-08/22/11.** They demonstrate a significant stability of an average price with a tendency to grow but a significant increase of standard deviation in price with subsequent flattening out.

| Trading days | Average price ($) | Standard deviation ($) | Trading volume (in 1000 shares) | Cumulative trading |
|---|---|---|---|---|
| 1 | 24.0617 | 0.145818 | 54,183 | 54,183 |
| 2 | 24.1733 | 0.190368 | 76,505 | 130,688 |
| 3 | 24.3891 | 0.316262 | 104,198 | 234,886 |
| 4 | 24.5429 | 0.447915 | 49,585 | 284,471 |
| 5 | 24.673 | 0.512251 | 52,847 | 337,318 |
| 6 | 24.737 | 0.536929 | 55,365 | 392,683 |
| 7 | 24.818 | 0.509597 | 64,419 | 457,102 |

[place Figure 3 approximately here]

To exclude the influence of the real macroeconomic events from short-horizon (1-10 days) VWAP distributions, we use the following method. At the first stage, we model the intraday NTSE-100 index. Simultaneously, we take the list of all substantial economic events in their intraday sequential order according to the WECO function of Bloomberg(C) and use Bloomberg-assigned weights and the news sign (positive-negative with respect to prediction by the analysts' community) and :

$$\vec{X}_t = \hat{B} r_t + \hat{C} F_t + \vec{\varepsilon}_t$$
$$\vec{X}' = (r, F)$$
$$\hat{B}' = (\sum_{j=1}^{5} b_j L^j, 0), \hat{C}' = (\sum_{m=1}^{3} f_m, f_4 \rho L)$$

(1)

Here, in Equation (1), $r_t$ are intraday index returns (taken with the standard 6 min. interval). $L$ is the one-period lag operator. Economic factors $f_m$, $m=1 \div 4$ are the coefficients measuring response of the index to the economic announcements $F$. The announcements are assigned signed numerical weights according to the Bloomberg(C) classification. Signs reflect the sign of surprise (positive or negative) with respect to the analysts' expectations. Also included as the explanatory variables their product (cross-influence) and integrated product with a weight $\rho^t$ for the lag $t$ put on the past events.[3] Thus constructed vector of economic news $F$ (850 announcements during 140 trading days) has approximately normal distribution.

The last factor reflects non-instant incorporation of economic events in the index price. Our estimates below (see Table 2) indicate that current economic news is fully incorporated into the index during 25-30 minutes of trading.

Explanatory power of this model is measured by OLS regression of calculated (Equation (1) without the error term) and the actual returns. Explanatory power is low ($R^2 \approx 1\%$) and we conclude that the bulk of intraday fluctuations of the stock prices within the bid-ask window (see Cao, Hensch and Wang, 2009, Fig. 1 for the sketch) is produced by microstructure effects. The fact that the slope of the OLS regression is close to unity and the intercept is close to zero suggests that the model of Equation (1) is correctly describing essential features of the intraday price process despite its low explanatory power.

We have tried several alternatives of the regression of Equation (1) explaining stock fluctuations by the intervening economic events. For instance, we tried to predict daily residuals of CAPM regression from the moving average of the strengths of the past announcements:

$$r_t = \alpha + \beta r_{Mt} + \delta \sum_{1}^{5} f_m \cdot h_{t-m} + \varepsilon_t$$

$$r_t = \alpha + \beta r_{Mt} + \delta \sum_{1}^{5} \tanh(f_m \cdot h_{t-m}) + \varepsilon_t$$

(1')

In the Equations (1'), $r_{Mt}$ is a return of a stock on some broad market index, $h_m$ are the measured strengths of the preceding announcements and $\delta$ is a scaling factor. Our two regressions: linear (first equation of (1')) and logistic (second equation) produce roughly similar results. The regressions of Equation (1') for the first 140 days preceding our measurement interval have similar predictive quality as the regression of Equation (1). The coefficients describing influence of past economic announcements have the same order of magnitude. Unlike the intraday model, the coefficients of the OLS comparison of the daily model with actual returns differ insignificantly from zero (see Table 2). Hence, we conclude that (1) intraday returns have significant dependence on the announcements of substantial economic news but they contribute relatively little to volatility and (2) these announcements are completely incorporated into the returns within one day and do not have any predictive value for the daily movements of the stock. Consequently, expected influence of the macroeconomic news on the shape of VWAP distributions from day to day is relatively small.

---
3   This is, of course, equivalent to the auxiliary AR(1) process for the factor $f_4$.

Table 2. **Intraday regressions of NTSE-100 for 19 trading days (07/25/11-08/18/11) with 6 minute interval according to Equation (1) and daily regressions for 140 trading days (01/04/11-08/17/11).** A) Parameter values for regression of Equations (1) is denoted as (1) and the Equation (1')—as (2). B) Relationship between the predictable and actual returns by OLS regression.

A)

| Estimation parameter | Magnitude (1) | Magnitude (2) |
|---|---|---|
| $b_1$ | -0.0218 | - |
| $b_2$ | 0.0214 | - |
| $b_3$ | -0.0122 | - |
| $b_4$ | 0.0166 | - |
| $b_5$ | 0.0181 | - |
| $f_1$ | -9.00E-006 | 1.64E-003 |
| $f_2$ | -4.60E-003 | -1.28E-003 |
| $f_3$ | 6.91E-003 | 1.04E-003 |
| $f_4$ | -5.54E-003 | -3.80E-004 |
| $f_5$ | - | 7.70E-004 |
| $\rho(\delta)$ | 0.2360 | 1.2033 |

B)

| Parameter | Regression (1) Magnitude (Std Err at 5%) | Regression (2) Magnitude (Std Err at 5%) |
|---|---|---|
| Intercept | -8.11E-05 (7.64E-05) | -2.1E-004 (1.61E-004) |
| Slope | 0.974 (0.222) | 0.019 (0.012) |
| MSE/SSE | 3.35E-04 (2.85E-03) | 9.42E-06 (3.63E-06) |
| $R^2_{,adj}$ | 1.30% | 1.14% |
| Regression valid at 5% | **Yes** | **No** |

[Place Figure 4 approximately here]

## 3. Aliasing of the VWAP distributions

At the second stage, we use method of aliasing borrowed from image processing techniques (Mallat, 1999). Namely, we use a Gaussian smoothing filter with the width corresponding to the standard deviation of predictable (past history and economic-defined) events as a smoothing kernel (see Tsay, 2002) for the empirical distributions both in state variable (price) and frequency domains. This intuitively corresponds to "averaging out" substantial economic events for the entire period.

$$F_j(p,\omega) = \sum_{p',\omega'} f_j(p-p', \omega-\omega') \hat{F}_j(p',\omega') \qquad (2)$$

In Equation (2), $\hat{F}_j(p,\omega) = P_{VWAP}(p) \cdot \varphi_{p \to \omega}[P_{VWAP}(p)]$ is an observed Volume-by-Price distribution for the $j$ integration days and $\varphi(.)$ is its Fourier transform. Furthermore, $f_j$ is a Gaussian aliasing kernel with the Gaussian width $\sigma_p$ in price and $\sigma_\omega$ frequency space (see Mallat (1999), Section 2.3.2) obtained from the first-stage regression. For the squares of Gaussian widths, we use sample estimators of the variances of the predictor of the (Equation 1) of the original signal and its Fourier transform, respectively. These sample estimators are scaled in proportion to the actual variance of the distribution of stock returns.

Intuitively, $F_j(p,\omega)$ is a distribution, for which actual macroeconomic events happening on observation days are replaced by their average intensity (volatility). The kernel $f_j$ has conventional diffusion scaling with the number of integration days:

$$f_j(p,\omega) = f_1\left(p\sqrt{j}, \frac{\omega}{\sqrt{j}}\right) \propto e^{-\frac{p^2}{2\sigma_p^2 j} - \frac{\omega^2 j}{2\sigma_\omega^2}}$$

The above procedure is somewhat similar to the one we used in our paper (Lerner, 2013) to separate noise, microstructure effects and economic events in intraday returns on NTSE-100 index.

### 4. Formulation of the LOB theory by Foucault, Kadan and Kandel

Foucault, Kadan and Kandel (2005) assume that a risky security is traded in a continuous double auction. Information is symmetric and all participants are liquidity traders, i.e. they trade independently of the market fundamentals. The only difference between traders is their tolerance for the speed of execution of their orders, quantified by the "patience" parameter, $0<\theta_P<1$, or the proportion of patient traders in the crowd.

In FKK, as well as in empirical reality, the number of trades matters much more than the volume (Glaser and Weber, 2007). In FKK, the buyer always follows the seller because all liquidity is supplied endogenously. Consecutive orders are numbered by an integer $j$, $j \in \{0,1,....s-1\}$, where $s$ is the length of a session. If the tick size is $\Delta$, then the updated prices will be:

$$P_{buy}(j) = a - \Delta \cdot j \qquad (3)$$
$$P_{sell}(j) = b + \Delta \cdot j$$

where $a$ and $b$, respectively, are the ask and bid prices in the beginning of the trading session. Equation (3) assumes that the price has been updated $j$ times to a moment at which the prices were observed.

In the FKK model there is a technical assumption that $a, b \in [A, B]$, where A and B are acceptable price limits. This assumption is not entirely unrealistic because in many markets there are circuit breakers preventing extreme price movements. For market orders, we automatically presume $j=0$.

The time for their execution of each order (a waiting time) is a random function $T(j)$. Traders optimize the waiting losses:

$$c_j = j_i - \delta_i \cdot T(j) \geq 0 \qquad (4)$$

where $i=I, P$ ("Impatient", "Patient") indicates the degree of immediacy (delay of execution) that each type of trader is willing to tolerate. The variables $\delta_i$ determine the potential loss expected by each type of trader for the unit time delay of the execution of their order. By construction $\delta_I > \delta_P > 0$. When, for a certain $j^*$, the inequality in Equation (4) reduces the equality of Equation (4) to zero, the corresponding price according to Equation (3) is called the reservation price and the time of execution the reservation time. The meaning of this equality is that the reservation price by definition is the price at which the trader is indifferent between the limit and the market order.

Further, FKK assume that a waiting time for the $j^{th}$ sell order, which follows the $j^{th}$ buy order, is distributed according to a Poisson distribution with the rate constant $\lambda$ having the unit of inverse time.

For the execution time, FKK derive an equation:

$$T(j) = \frac{\alpha(j)}{\lambda} + \sum_{k=1}^{j-1} \alpha_k(j)[1/\lambda + T(k) + T(j)] \qquad (5)$$

In Equation (5), $\alpha_k(j)$ is the probability of a limit order with the spread $k$ on the $j^{th}$ step. The meaning of Equation (5) is as follows. The first term indicates the probability that all orders up to $(j-1)^{th}$ were market orders. The expected delay for the execution of the limit order with the observed spread $k$ is larger than the expected delay of the market order $1/\lambda$ by the sum of the expected delay of the trader of the opposite kind (buyer for seller and seller for buyer) $T(k)$ and the expected delay of the trader of the same kind $T(j)$. Furthermore, by the classic rule for probabilities:

$$\sum_{k=1}^{j-1} \alpha_k(j) = 1 \tag{6}$$

and the Equation for the *T(j)* acquires the form:

$$T(j) = \frac{1}{\alpha_0(j)} [1/\lambda + \sum_{k=1}^{j-1} \alpha_k(j) T(k)] \tag{7}$$

Then, FKK proves that the patient trader *i=P* facing the spread within the limits <$n_h +1, n_h$>, for *h=1, ... ,q-1* where $n_1 = j'_P$ is the reservation spread for a patient trader, and the trader numbered $n_q$=K≡a-b submits a limit order at her reservation price j=$n_h$·Δ. Hence, the equilibrium distribution of delay times follows from Equation (7) as:

$$T(n_h) = \frac{1}{\lambda} \cdot [1 + 2 \sum_{k=1}^{h-1} (\frac{\theta_P}{1-\theta_P})^k] \tag{8}$$

where *h=2, ... ,q-1*. This equation relies on the observation that, on their way to *h*'s trade, the patient trader randomly met patient and impatient traders in direct proportion of their occurrence $\theta_P$ and (1 – $\theta_P$) in the sample.

Because of Equation (4), the distribution of delay times is proportional to the cost function for the traders:

$$c_j(n_h) = j_i - \delta_i \cdot T(n_h) \tag{9}$$

Expected prices in the efficient market are expressed through the proportion of patient traders $\theta_P$ as follows:

$$\begin{cases} P_{buy} = a - \Delta \cdot \theta_P \cdot j_P - \Delta \cdot (1-\theta_P) \cdot j_i \\ P_{sell} = a + \Delta \cdot \theta_P \cdot j_P + \Delta \cdot (1-\theta_P) \cdot j_i \end{cases} \tag{10}$$

Equation (10) reflects the fact that, due to the presence of two groups of traders, selling pressure reduces the ask price and raises the bid price, in equal amounts. The mid-price, in the original formulation of the FKK model, stays the same and we continue this convention for clarity, though this assumption can be modified for enhanced realism.

For a very large number of traders/adjustment steps $h \to \infty$ (very liquid stock), distribution of waiting times (Equation (8)) tends to constant, which depends only on the fraction of impatient traders:

$$T(n_h) \to T_\infty = \frac{1}{\lambda(1-2\theta_P)}, \quad T_\infty^{-1} = \lambda \cdot (1-2\theta_P) \tag{11}$$

We note that, while the process of quotes update is not a Markov process (waiting time depends on prehistory), asymptotically, for the large number of traders it is a Markov process. We shall exploit this property in the next section.

In the absence of intervening economic events, the expected mid-price always stays the same. Equation (8) and its derivatives such as (11) cannot easily be compared with empirical data, at least, not unless all limit orders with time stamps are known. Therefore, we propose a theoretical setting, which can potentially lead to observable quantities. Namely, suppose that each trading session has a fixed number of patient and impatient traders but that what we observe in the market is a representative number of these trading sessions. In that case averaging produces statistical distribution of the patient/impatient traders in time, which we identify with VWAP.

## 5. PDE governing traders' distribution

To express the FKK theory in a potentially verifiable form, we assume that the quotes are governed by a two-state Markov stochastic process $\theta_t$ instead of a static parameter $\theta_P$. This process indicates a kind of trader (patient or impatient) executing on the market at a given time. Two values of this process are equal to $\theta_P$ and $(1-\theta_P)$ and are independent from the market noise. The transition probability matrix describes the following situation:

| Departure of patient traders from patient state | Arrival of impatient traders in place of patient ones |
|---|---|
| Arrival of patient traders in place of impatient ones | Departure of impatient traders from impatient state |

Markov dynamics are fully determined by the transition matrix between states. In our case, following the usual assumptions, we choose the transition matrix consistent with Equation (11) in its simplest form:

$$\hat{T} = \begin{pmatrix} -\lambda & \lambda \\ \lambda & -\lambda \end{pmatrix} \quad (12)$$

Then, the probability of the patient trader continuing to wait $\vartheta = P[\theta_t \leq \vartheta]$ obeys the Equation:

$$\frac{d\vartheta}{dt} = \lambda(1 - 2\vartheta) \quad (13)$$

This Equation indicates that with a rate $\lambda(1-2\vartheta)$, the patient trader is being replaced by an impatient trader and vice versa. The given form of the transition matrix (13) is not unique, on the contrary, there is an infinite number of Markov transition matrices corresponding to the asymptotic law of Equation (9) but we shall use the 2×2 transition matrix for simplicity. Equation (13) is deterministic, albeit with the stochastic initial condition—initial distribution of $\theta_P$—and can be associated with the state equation of the filtering problem. The meaning of Equation (13) is that asymptotically, in the limit of the large number of trades, a convergence to the true price is negative-exponential.

One cannot observe real-time proportion of patient and impatient traders. We assume that we can glean it with some inevitable stochastic error through the volume-at-price distribution (see the next section for explanation):

$$d\omega = \vartheta \, d\tau + \sigma \, dW_\tau \quad (14)$$

In Equation (14), as usual, $W_\tau$ is a stochastic noise process assuming a continuous Gaussian Markov process for analytic tractability,[4] the variable $\tau = T - t$. Equation (14) performs a role of an observation equation in the filtering problem.

---

4  Ait-Sahalia and Jacod (2012) suggest that price jumps play a fundamental role in time-series statistics of asset returns. Microstructure noise in continuous time is much more likely to resemble white noise. We shall partially correct the restrictions of our assumption for the microstructure noise by assuming Poisson statistics for the short term fluctuations of quotes arrival (see Section 5).

As the state variable we chose the conditional expectation that at the inverse time $\tau = T-t$, a randomly chosen trader from the sample of traders active during the interval [t,t+dτ] belongs to either the "I" or "P" group: $\omega(\tau) = E[\vartheta = \theta_P | F_\tau^W]$. We need only one variable because of the dichotomy of patient and impatient traders in the FKK model. The set $F_\tau^W$ refers to the microstructure noise. To derive a dynamic equation for the state variable, we replaced discretized spreads of Equation (3) by a random walk with an instant coordinate $\vartheta$ corresponding to $n=n_h$.

Equation (11) expresses delay of execution through a fraction of patient traders, $\theta_P$. Inverse to the time $T_\infty$ is the average rate at which orders get executed in the limit of a large number of trades. Impatient traders, for instance buyers, want to execute their successive trades during a characteristic time $T_1=1/\lambda$. However, they might not find equally impatient seller and have to wait until a patient seller arrives. Yet, patient traders have a distribution of execution times between $T_1=1/\lambda$ and $T_\infty(\theta_P) \to \infty$, for $\theta_P \to 1/2$.

It turns out that starting from Equation (13) we can derive, in the continuous limit, a PDE governing the distribution function of ω (see Appendix A). Equation (10) permits us to express the price distribution from this function.

A conventional Kalman-Bucy filtering procedure (for instance, Liptser and Shiryayev, 1977, Theorem 9.1) leads to a backward Kolmogorov equation for the distribution function (see Appendix A):

$$\frac{\partial \pi(\omega,\tau)}{\partial \tau} = -\lambda[(1-2\omega) - \frac{\mu}{\lambda}\omega(1-\omega)]\frac{\partial}{\partial \omega}\pi[\omega,\tau] \\ + \frac{\lambda^2 \sigma^2}{2}(\omega^2(1-\omega)^2 \frac{\partial^2}{\partial \omega^2}\pi[\omega,\tau]) \qquad (15)$$

The analogues of the evolution Equation (15) and the Hermitian conjugate of the evolution operator (forward Kolmogorov equation) appear in many contexts (except the already quoted Liptser and Shiryayev, 1977, see, for instance Bharucha-Reid, 1960, Chapter 4.5 and Çetin and Vershuere, 2010). We shall discuss biological analogies and interpretations below, in Section 6.

The terminal condition for $t=T$ for this Equation can, for instance, approximate a stationary solution of Equation (13) (Appendix A). One of the simplest approximations of bimodal distribution is:

$$P(\omega, n_h \to \infty) = A \cdot \delta(\omega - \theta_1) + B \cdot \delta(\theta_2 - \omega) \tag{16}$$

where $\theta_1$ and $\theta_2$ are the lower and upper limits on the proportion of patient traders. If $A+B=1$, the constants $A$ and $B$ can be identified as the proportion of patient buyers and patient sellers, respectively.

The terminal condition of Equation (16) simply means that for a sufficiently long trading session all patient traders-buyers concentrate near a bid price and patient traders-sellers—near an ask price. Note that the asymptotic value of the limiting proportion of patient traders, $\theta_P$, cannot exceed ½ (and not one!) because the series for $T(n_h)$ diverges for a large number of executed trades when the proportion of patient traders exceeds one half.

Discontinuous terminal conditions of Equation (16) can pose a difficulty for numerical modeling. In our simulation we replace them with a suitable continuous approximation of a δ-function, for instance, a narrow Gaussian shape. The question of the boundary conditions requires additional considerations, which are presented in Appendix B.

## 6. Price distribution in the dynamic LOB model

Equation (15) with appropriate terminal and boundary conditions gives a probability distribution at a given time $\tau$ for a fraction of patient traders during the trading session. This equation is, in principle, not very different from the celebrated Black-Scholes equation. However, unlike the Black-Scholes where the state variable is the stock price $S$, here, the state variable is the probability of the next trader being patient or impatient. This is not an empirically observable quantity.

To compare the computed distribution of patient traders with any empirical price distribution, one needs to use Equation (3) or (10) and the Bayes formula. Of course, this price distribution depends on the lower and upper limits of prices, which are the results of economic fundamentals. In our presentation, for clarity, we consider that the economic fundamentals change infrequently. For practical calibration, this assumption might be revised.

The Bayesian method used in this paper is similar to the one used by Merton (Merton, 1973). The foundations of Merton's method using modern stochastic calculus techniques can be found in Jeanblanc, Yor and Chesney (2009) and the references therein.

We define a price distribution for bids in a real time $t$ as an expectation that at a time $t$ a trade will be executed at a reservation price $P(j')$, compared with Equations (3) and (4):

$$\pi_a(p,t) = E_t^X[p = a - P_{buy}(j') | \theta_P = \omega(t)] \qquad (17)$$

Similarly, the price distribution for the ask prices can be defined as:

$$\pi_b(p,t) = E_t^X[p = P_{sell} - b(j') | \theta_P = \omega(t)] \qquad (18)$$

In Equations (17) and (18), $X$ is a symbolic notation for the terminal and boundary conditions in Equation (15) and $j'$ is the reservation spread.

In the FKK framework, the bid prices are always adjusted upwards (Equation (3)) and the ask prices are always adjusted downwards with respect to the trading limits $a$ and $b$. These limits depend on economic fundamentals, which we consider as being infrequently revised, which is the practical case for almost all stocks. Even for the most liquid companies big news does not arrive every day. As one can glean from Equation (10)—for instance from the fact that in FKK, mid-price is unchanged during trading— these distributions are identical, so we omit the upper and lower indexes.

The function is obviously a probability distribution because (a) the expected price adjustment $p$ is non-negative—nobody would bid at a price above an already available selling price or ask below a current buy price—and (b), it is normalized to unity:

$$\int_0^\infty \pi(p,t)\,dp = 1 \qquad (19)$$

Normalization follows from the simple fact that for any moment in real time we expect the price change $p>0$ to be equal to *something*. Actual limits of integration consistent with the FKK in the integral of Equation (19) lie between *a* and *b*—the starting prices at the beginning of the stylized trading session.

Further, we can use the Bayes formula and Equation (3) to derive the observable price distribution in real time $P_{obs}(p,\tau)$:

$$P_{obs}(p\equiv\delta\cdot\omega/\Delta,\tau)=\sum_{i=1}^{\infty}\omega\pi(\omega,\tau=\lambda i)\cdot P_{n,P}(i,t) \\ +\sum_{i=1}^{\infty}(1-\omega)\pi(\omega,\tau=\lambda i)\cdot P_{n,i}(i,t) \quad (20)$$

where $P_{n,P}(i,t)=P_{patient}(t|n_h=i)$, $P_{n,i}(i,t)=P_{impatient}(t|n_h=i)$ are the prior price distributions conditional on the average number of price adjustments at a time $t<T$. For convenience, in Equation (20), we rescaled price into units of $\Delta$—the jump size—in Equation (3). In writing down Equation (19) we assumed that the distributions of the price adjustments are independently normalized on unity. In Equation (20), $\pi(\omega,\tau)$ is the result of numerical integration of Equation (15).

Equation (20) gives a price distribution in a parametric form dependent on a fraction of impatient traders. To compare it with observable Volume-at-Price distributions, one needs to complement the Bayesian formula (20) with an appropriate prior for the state variable.

In our case, prior distributions of impatient and patient traders are assumed to be Poisson distributions with rate constants equal to $T_1^{-1}=\lambda$ and $T_\infty^{-1}=\lambda(1-2\vartheta)$, respectively.[5] This is, of course, a crude approximation but I expect that the difference with the true distribution for a few days would be small.

In Fig. 4, I present the simulations of the trading market model of Equations (15) and (19). They reproduce the main qualitative feature of the empirical Volume-at-Price distributions: nearly bell-shaped distribution for a small quote integration time, a shapeless form for a few days and progression to a distinct bimodal distribution with modes near the upper and lower limits of trading for the period. The existence of the two humps (otherwise known as bimodal structure of the distribution) is not surprising—it is an artifact of the terminal condition of Equation (16)—but Equation (15) demonstrates an evolution from this distribution into a familiar bell-shaped curve.

---

5  For modeling using Poisson distributions, see e.g. Gourieroux and Jasiak (2001) or Chernobai, Rachev and Fabozzi (2007).

The simulations were performed for relatively arbitrary values of the parameters (unit of trading time, duration of the period, proportions of patient buyers and sellers and maxima for terminal distribution of patient buyers and sellers, respectively) $\lambda=0.25$, $T=10$, $A=B$ and $\theta_1 \approx 0.13$, $\theta_2 \approx 0.38$. The moments of the simulated distribution behave approximately as the moments of the real distribution integrated for 10 days (Table 3).

Table 3. **First two moments of the analytical distributions**. Numerical analytic distributions were approximately rescaled to conform to the scale of the empirical distributions of Table 1. Numerical data, corresponding to the values of parameters (unit of trading time, duration of the period, proportions of patient buyers and sellers and maxima for terminal distribution of patient buyers and sellers, respectively) $\lambda=0.5$, $T=25$, $A=B$ and $\theta_1 \approx 0.13$, $\theta_2 \approx 0.38$, $\theta_P=0.1$ were not calibrated to the empirical data and are given here for illustrative purposes.

| Rounded-off Time (days) | Average price ($) | Standard deviation, ($) |
|---|---|---|
| 2 | 24.19 | 0.143 |
| 3 | 24.25 | 0.170 |
| 6 | 24.6 | 0.490 |
| 7 | 24.96 | 0.512 |

[place Figure 4 approximately here]

7. **Biological analogues of the competition between patient and impatient traders**

Many biological/ecological systems express features similar to those of the dynamic LOB evolution. Their review can shed light on the intuitive nature of the results of previous sections. In the problem of competition between two alleles of the same species, if an appropriate mutation increases the longevity of the fertile period of an organism, it increases intergenerational representation of a longer-living species and thus assures the dominance of this species in a distant future. In the beginning, there will be a random distribution of alleles but as time goes by, only the allele with a useful mutation survives. In the end, the age distribution will be dominated by the longer-living, i.e. "patient" species. This view is a variation of the well-studied Kimura (1954) model in mathematical biology. (Bharucha-Reid, 1960, Chapter 4.5) Different versions of the Kimura model are possible and were developed in the context of mathematical genetics.

If, for instance, instead of increased survival, we assume that mutation increases only the longevity (but not, for instance, the average fertility rate) of the mutated species, the model will be described, instead by the Kimura equation, by an equation very close to the Equation (13).

In application to finance, traders who have a relatively small penalty for waiting for a more advantageous price benefit more at the expense of traders who value immediacy. Eventually, however, with the depletion of the pool of impatient traders, patient traders are forced to trade with one another and the VWAP pattern settles into the buyers concentrating near the selling price and sellers near the buying price with occasional clearance happening when a surviving impatient trader completes a deal.

One of the more profound analogies of this situation has been modeled by the Nobel Prize Winner in biology Edward O. Wilson in his 1980 paper with Lumsden (Lumsden and Wilson, 1980). In that paper, the authors postulated the existence of evenly distributed "cultural" patterns between societies, which can be taught and learned. Conformity to these patterns, called "trend-watching" , improves procreative chances of an individual (similar to competition for the best price of execution in the financial literature (Cont, Stoikov and Talreja, 2008) resulting in a broadened distribution of genetic markers. Through generations, small initial differences in learning bias can be significantly amplified. The affinity of Lumsden-Wilson simulated distributions and our simulations (compare Figs. 3 and 4 in Lumsden-Wilson, 1980 and our Fig. 4) in the case they called "non-saturable trend-watching" is striking. The authors modeled the preference by an exponential "assimilation function" while its analog in our treatment is parabolic.

Lumsden and Wilson were concentrating on a stationary distribution (Lumsden and Wilson, 1980, Eq. (5)), which can be easily inferred from a forward Fokker-Plank equation:

$$P(\theta) = C \left(\frac{\theta}{1-\theta}\right)^{\alpha} e^{\frac{1}{\lambda \sigma^2}(1/\theta + 1/(1-\theta))} \quad (19)$$

where $\alpha = 2\mu/\lambda\sigma^2$ and C is a normalization constant. In our case, this solution does not play an important role because in real trading markets, the equilibrium is systematically violated by the arrival of substantive news.

**8. Note on the predictable size of the effect**

We used a very liquid stock (Microsoft) for our illustration. Furthermore, only a 10-day integration horizon was available for the VWAP function in the 2011 version of Bloomberg(C), which limited our lookback capacity to seven trading days.[6] Thus, a size of infrastructure effects observed in our example is likely to be underestimated with respect to longer samples and less liquid securities.

A natural estimate of the trading loss of impatient traders as a group to the patient traders would be the standard deviation of VWAP for the unimodal distribution (~$0.15 in our example) or half of the standard deviation—an approximate width of a single peak—$0.25 for the bimodal distribution. Given that the typical price of Microsoft stock in 2011 hovered around $25, a trading loss because of immediacy concerns can be crudely estimated as -0.7%-1% per day.

Accrual of this loss throughout longer periods is harder to estimate because during longer periods the same agent can appear on the patient as well as the impatient side. If we identify the accumulation of transaction cost losses with the cost of maintenance of the portfolios, i.e., this cost scales approximately as a square root of the number of trading periods (Bollen, Whaley and Smith, 2004, also Lerner, 2009, Chapter 2.2), then during one month assumed as 22 trading days:

---

6  Intraday, full-quote VWAP was not available for all stocks in the 2011 version of Bloomberg(C) subscribed by Cornell University and we also needed significant quote statistics for smooth, quasi-continuous distributions without holes.

$$L \simeq -70 \div 100 \, bp \sqrt{22} \simeq -300 \div 470 \, bp$$

This estimate demonstrates that even for a very liquid stock, trading losses due to microstructure frictions can quite substantially eat into the return if trades are scheduled unwisely. Here, I must provide a cautionary note that the above considerations do not contradict profitability of the "front-running" strategies such as detected by Da, Gao and Jagannathan (Da, Gao and Jagannathan 2011, see also preface to Lerner, 2009 with the description of some of the abusive methods of front-running). It relies on the speed of reaction to real, imaginary or manufactured economic events and the trader can benefit from even a very unfavorable execution price if the movement of the stock price in response to the information event is significant.

**Conclusion**

In this paper, I develop a plausible argument that a relatively parsimonious extension of the Foucault-Kadan-Kandel (FKK, 2005) theory can explain some qualitative features of Volume-at-Price distribution for stocks with a large depth. I demonstrate that asymptotically, in the limit of a large trade size, FKK leads to a relatively simple PDE, which can be solved numerically.

Simulations demonstrate the same qualitative feature as the real VWAP plots, namely, the spread shrinks even without intervening economic news, which is absent in our stylized model. However, the standard deviation of price distribution grows in the beginning of the process.

Because the model is symmetric with respect to buying and selling but its buying and selling distributions can be independently calibrated with the available market data (see Fig. 1), one can hope to quantify notions of "buying" and "selling pressure" in the framework of the presented model. Quantitative comparisons are currently a work in progress. The impediment is that the calibration of parameters in a Poisson-weighted family of solutions of partial differential equations is not a straightforward econometric problem.

We estimated the size of the expected effect. If we suppose that in the end of the trading period, all patient sellers concentrate near the ask price and all patient buyers concentrate near the bid price, then the size of the effect is roughly a half of the bid-ask spread for the regular days without much imbalance for buy and sell orders. Daily accumulation of these losses can reduce the performance of portfolios quite substantially.

Even for the most liquid stocks, such as AOL, Microsoft and 3M, this effect is on the order of magnitude of a dozen-or-so cents for the share price of several tens of dollars. That is, it is on the order of a few percent, which results in several million a day transferred from impatient to patient traders for a typical turnover of a highly liquid S&P or Nasdaq stock.

Appendix A. **Filtering equation for the observation of quote-updating process**

A derivation of the PDE for the observation process described in Section 3 is performed by the standard techniques of Kalman-Bucy filtering. In the treatment below, we closely follow Liptser and Shiryaev (1977). Let $(\vartheta, \omega)$ be the random process introduced in Section 3, where $\vartheta$ is a "true" traders' arrival process and $\omega$ is the process as it appears to the observer. Then, one can define a conditional probability

$$\omega_\beta(t) = P[\theta_t = \beta | F_t^W], \quad \beta = i, p \tag{A.1}$$

which is the best guess of the type of trader based on the previous trading information.

Then, under some generic conditions on the nature of probability space, Liptser and Shiryaev prove the following theorem. If the processes $(\vartheta, \omega)$ are diffusions driven by the Brownian noise term, then the probability $\omega_\beta(t)$ obeys the equation of diffusion

$$d\omega_\beta(t) = a(\omega_\beta, t) dt + b(\omega_\beta, t) d\bar{W}_t \tag{A.2}$$

where the coefficients a and b are given by the following equations:

$$a(\omega, t) = \sum_{\alpha \in E} \lambda_\alpha \omega_\alpha(t) \tag{A.3}$$

$$b(\omega, t) = \sigma[A_t(\omega, \xi) - \sum_{\alpha \in E} A_t(\alpha, \xi) \omega_\alpha(t)] \tag{A.4}$$

The terms $\lambda_\alpha$ in Equation (A.3) are the elements of the Markov transition matrix (Equation 12 of the main text). $A_t$ is the evolution rate operator from Equation (9). The Brownian motion $\bar{W}_t$ is a Wiener process for innovations, the exact nature of which is not relevant for us now (e.g. Liptser and Shiryaev, 1977, Theorem 9.1). An explicit expression for $\bar{W}_t$ is given in Theorem 8.1, ibid. In (A.2), we presume that the number of discrete values of our process $E$, is at least countable (it is 2—patient and impatient—in our simplified formulation). Because we have only two states, $\omega_p = 1 - \omega_i = \omega$ and we can take the probability of patient trader as the only state variable. The differentials of these variables are related as $d\omega_i = -d\omega_p$. Using (A.3) and (A.4) and Equations (9-11) of the main paper we get:

$$a(\omega, t) = -\lambda \omega + \lambda(1-\omega) = \lambda(1-2\omega) \tag{A.5}$$

and

$$b(\omega, t) = \lambda \omega \sigma (1-\omega) \tag{A.6}$$

The resulting diffusion takes a form:

$$d\omega(t) = \lambda(1-2\omega)dt + \lambda\sigma\omega(1-\omega)d\bar{W}_t \qquad (A.7)$$

However, Equation (A.3) reflects the situation where the choice between immediate execution and waiting is risk-neutral. But there is no logic in assuming that, given that the order of execution is not tradable between parties.[7]

Because of that, $\bar{W}_t$ is no longer a Brownian motion under $P[.|F_t^W]$. Generally, its form can be quite arbitrary. However, if we assume that the distribution of $\vartheta_T$ at the final moment of the trading session is known, in an extended algebra $F^{W \vee \theta_T}$ there exists a numerical constant µ such that a transformed process is a Brownian motion under $P[.|F_t^{W \vee \theta_T}]$:

$$d\bar{W}_t \to d\bar{\bar{W}}_t - \mu\omega(1-\omega)dt \qquad (A.8)$$

(see, e.g. the treatment in Çetin and Vershuere (2010)). Constant µ has the intuitive meaning of a market price of delay risk. The final equation for the diffusion becomes:

$$d\omega_\beta(t) = \lambda\left[(1-2\omega) - \frac{\mu}{\lambda}\omega(1-\omega)\right]dt + \lambda\sigma\omega(1-\omega)d\bar{\bar{W}}_t \qquad (A.9)$$

Equation (A.9) can be rendered in the form of the forward Kolmogorov diffusion for $\pi(\bar{\omega}, t) = dP(\omega_t \leq \bar{\omega})/d\bar{\omega}$ — the probability distribution for the fraction of patient vs. impatient traders:

$$\frac{\partial \pi(\bar{\omega}, t)}{\partial t} = \lambda \frac{\partial}{\partial \bar{\omega}}\left[(1-2\bar{\omega}) - \frac{\mu}{\lambda}\bar{\omega}(1-\bar{\omega})\right]\pi(\bar{\omega}, t) + \frac{\lambda^2\sigma^2}{2}\frac{\partial^2}{\partial \bar{\omega}^2}\left[\bar{\omega}^2(1-\bar{\omega})^2\pi(\bar{\omega}, t)\right] \qquad (A.10)$$

We use the backward form of Equation (A.10), $\tau = T-t$ (Equation (13) of the main text) in our simulations and the forward form in Appendix B. Elsewhere, for brevity we omit the bar over the argument, because a subtle difference between $\omega_t \equiv \omega(\tau)$ —the state variable and $\bar{\omega}$ —the parameter in distribution function, cannot cause confusion in other contexts.

---

7 Of course, this assumption would be violated if high-frequency traders are allowed to bid higher prices for speed of the execution. Consequences of more complete high-frequency markets are unclear at this point but are certainly worth investigating.

Appendix B. **Boundary conditions for Kolmogorov-Fokker-Planck (KFP) equation**

Equation (13) has the diffusion-dissipative type like all the Black Scholes-type (or Fokker-Kolmogorov) equations. Because of that it is irreversible in time. In particular, only the forward equation is likely to preserve the probability normalization to unity.

We *postulate* that the boundary conditions for the backward (i.e. standard for finance) PDE must preserve probability for the conjugate forward equation. This postulate is not a consequence of any mathematical theory but it is consistent with common sense. In particular, it is a direct analog of the smooth-pasting condition in the endogenous default models in credit risk analysis (Mella-Barral and Perraudin, 1997).

Hence, the problem of probability conservation and the boundary conditions must be investigated for the forward KFP. We assume homogeneous boundary conditions of a Dirichlet or Neumann type.

The forward version of the Equation (15) is as follows:

$$\frac{\partial P(\omega,\tau)}{\partial \tau} = -\lambda \frac{\partial}{\partial \omega}\left((1-2\omega) - \frac{\mu}{\lambda}\omega(1-\omega)P[\omega,\tau]\right) + \frac{\lambda^2 \sigma^2}{2}\frac{\partial^2}{\partial \omega^2}(\omega^2(1-\omega)^2 P[\omega,\tau]) \quad \text{(B.1)}$$

in the Fick's (Klages, Radons and Sokolov, 2008) form:

$$\frac{\partial P(\omega,\tau)}{\partial \tau} = -\frac{\partial}{\partial \omega} j(\omega,\tau) \quad \text{(B.2)}$$

where the probability current is equal to

$$j(\omega,\tau) = -\lambda\left((1-2\omega) - \mu\frac{\omega}{\lambda}(1-\omega)P[\omega,\tau]\right) + \frac{\lambda^2 \sigma^2}{2}\frac{\partial}{\partial \omega}(\omega^2(1-\omega)^2 P[\omega,\tau]) \quad \text{(B.3)}$$

The conservation of the probability condition:

$$\int_0^{1/2} P(\omega,\tau) d\omega = 1 \quad \text{(B.4)}$$

can be differentiated in $\tau$ to yield:

$$\int_0^{1/2} \frac{\partial P(\omega,\tau)}{\partial \tau} d\omega = -\int_0^{1/2} \frac{\partial j(\omega,\tau)}{\partial \omega} d\omega = j(0,\tau) - j(1/2,\tau) = 0 \qquad (B.5)$$

Using Equation (B.3) to express the probability current on the boundaries through the probability, we arrive at the expression:

$$P(0,\tau) + \frac{\lambda \sigma^2}{8} \frac{\partial}{\partial \omega} P(\omega=1/2,\tau) = 0 \qquad (B.6)$$

This equation connects the probability distribution on the lower boundary $\theta_P=0$ with the first derivative of the probability on the upper boundary $\theta_P=1/2$. Assuming that if only impatient traders are present in the market, a boundary condition reads as:

$$P(0,\tau) = 0 \qquad (B.7)$$

The intuitive meaning of the Equation (B.7) is that, in the beginning of trading session, the numbers of impatient traders completely dominate the patient trading and only though continuing trading the presence of patient traders becomes visible. Then, equation (B.7) immediately gives:

$$\frac{\partial}{\partial \omega} P(\omega=1/2,\tau) = 0 \qquad (B.8)$$

The last equation provides a condition on the spatial derivative of the probability distribution on the upper boundary. Another assumption fixing the same boundary conditions would be that Equation (B.6) must hold for arbitrary $\lambda$.

**Figure Captions**

a) [Figure 1a]
b) [Figure 1b]

Fig.1 Evolution of Volume-at-Price distributions with sampling days. a) Volume-at-price distribution integrated for one day (Aug. 22, 2011) for the Microsoft stock. Distribution is approximately bell-shaped, b) VWAP distribution of AOL orders for the same day (blue) approximated by lognormal distribution (red).

a) [Figure 2a]

b) [Figure 2b]

Fig. 2 a) Integrated volume-at-price distribution of the Microsoft stock for five days (08/17/11-08/22/11). The distribution is extended along the price axis. Maximum is poorly discernible. b) Integrated VWAP for the MMM stock for the same period (red). Two overlapping peaks start to resolve (red). Approximation by two Gaussians is added as a guide (blue).

Fig. 3. a) Microsoft VWAP for the ten days (08/11/11-08/22/11). Ten days was a practical limit of quote collection allowed by the Bloomberg© system in 2011. We observe that the distribution has a tendency to concentrate in two peaks—near the minimum and maximum price for an observed period, b) De-trended and filtered VWAP distributions in accordance with precepts of the Section 4.

a) [Figure 4a]

b) [Figure 4b]

c) [Figure 4c]

Fig. 4 a) The solution of Equation (13) of the main text. Simulated VWAP distribution is plotted as a function of normalized price and time. In the beginning of time, there is unimodal distribution, which converges to a terminal bimodal distribution. b) Weighted $\theta$-distribution (Equation 18), which we identify with the Volume-at-Price distribution. c) Simulated price distribution in arbitrary units as a function of time. Small t=3 (short times of integration) displays almost a bell-shaped form (red curve); compare to Fig. 1a, for the medium $\tau$=22, the structure is relatively absent (green curve; compare to Fig. 1b), for the long integration time, t=25 the structure converges to bimodal shape similar to the real quotes in Fig. 2c. The scale for the intermediate time is enhanced fivefold; the plot for the large time enhanced 10 times for clarity.

# References

Aït-Sahalia, Y., J. Fan and Y. Li (2013) The Leverage Effect Puzzle: Disentangling Source of Bias at High Frequency at High Frequency, *Journal of Financial Economics*, 109, 224-249.

Aït-Sahalia, Y. and J. Jacod (2012) Analyzing the Spectrum of Asset Returns and Volatility Components in High-Frequency Data, *Journal of Economic Literature*, 50(4), 1007-1050.

Angel, J., L. Harris and C. S. Spatt (2010), Equity trading in the 21st Century, Marshall School of Business Working Paper No. FBE 09-10.

Barclay, M. and T. Henderschott (2004), Liquidity externalities and adverse selection: Evidence from trading after hours, *Journal of Finance*, 59, 681-710.

Bollen, N.P., Smith, T., and Whaley, R.E. (2004). Modeling the Bid/Ask Spread: Measuring the inventory-holding premium, *Journal of Financial Economics*, 72, 97-141.

Bharucha-Reid, A. T. (1960), *Elements of the Theory of Markov Processes and their Applications*, McGraw Hill: New York, NY.

Chernobai, A. S., S. Rachev and F. Fabozzi (2007) *Operational Risk: A Guide to Basel II Capital Requirements, Models, and Analysis*, Frank J. Fabozzi Series, Vol. 152.

Çetin U. and M. Verschuere, (2010) Pricing and hedging in carbon emissions markets, *International Journal of Theoretical and Applied Finance*, 12(7), 949-967.

Cont, R., S. Stoikov and R. Talreja (2008) A stochastic model of order book dynamics, Columbia University working paper, www.ssrn.com/abstract=1273160.

Da, Z., P. Gao and R. Jagganathan (2011) Impatient Trading, Liquidity Provision and Stock Selection in Mutual Funds, *Review of Financial Studies*, 24(3), 675-720.

De Jong, F. and B. Rindi (2009) *The Microstructure of Financial Markets*. Cambridge University Press: Cambridge, UK.

Easley D., M. Lopez de Prado and M. O'Hara (2011), The Microstructure of the 'Flash Crash': Flow Toxicity, Liquidity Crashes and the Probability of Informed Trading *The Journal of Portfolio Management*, Vol. 37, No. 2, pp. 118-128.

Foucault, T., O. Kadan and E. Kandel (2005), Limit Order Book as a Market for Liquidity, *Review of Financial Studies*, 4, 1171-1217.

Hasbrouck, J. (2007) *Empirical Market Microstructure*. Oxford University Press: New York, NY.

Harris, L. (2003) *Trading and Exchanges: Market Microstructure for Practitioners*. Oxford University Press: New York, NY.

Henderschott, T. and C. M. Jones (2005), Island goes dark: Transparency, fragmentation and regulation, *Review of Financial Studies*, 18, 743-793.

Glaser, M. and M. Weber (2007), Overconfidence and trading volume, *Geneva Risk and Insurance*, 32, 1-36.

Gourieroux, C. and J. Jasiak (2001), *Financial Econometrics: Problems, Models, and Methods*, Princeton University Press, Princeton: NJ.

Harris, L. E. (1998), Optimal Dynamic Order Submission Strategies in Some Trading Problems, *Financial Markets, Institutions and Instruments*, 7, 1-76.

Jeanblanc, M., M. Yor and M. Chesney (2009), *Mathematical Methods for Financial Markets*. Springer: Heidelberg, Germany.

Jovanovic, B. and A. Menkveld (2011), Middlemen in limit order markets, New York University working paper, ssrn.com/abstract_id= 1624329.

Klages, R., G. Radons and I. M. Sokolov (2008) *Anomalous Transport*, Wiley: Weinheim, Germany.

Lerner, P. (2009), *Microstructure and Noise in Financial Markets*. VDM Verlag: Saarbrucken, Germany.

Liptser, R. S. and A. N. Shiryayev (1987), *Statistics of Random Processes*, Vol. 1 and 2. Springer-Verlag: New York, NY.

Mella-Barral, P. and W. Perraudin (1997) Strategic Debt Service, *J. Finance*, Vol. 50(2), 531-556.

Menkveld, A. J. (2010), High frequency traders and the New-Market makers, VU University Amsterdam, ssrn.com/abstract_id=1722924.

Merton, R. C. (1973), Theory of rational option pricing, *Bell Journal of Economics and Management Science*, 4(1), 141-183.

Obizhaeva, A., J. Wang (2005) Optimal trading strategy and supply/demand dynamics, NBER Working Paper No. 11144.

O' Hara, M. (1995), *Market Microstructure Theory*, Oxford University Press: Blackwell, UK.

Skjeltorp, J., E. Sojli and W. Wah Tam (2012) Identifying Cross-Sided Liquidity Externalities, Erasmus University working paper, www.ssrn.com/abstract=2026593.

Sornette, D. and S. von der Becke (2011), Crashes and High Frequency Trading, *Swiss Finance Institute series* № 11-63.

Switzer, L. and H. Fan (2007), The Transaction Costs of Risk Management vs. Speculation in an Electronic Trading Environment: Evidence from the Montreal Exchange, *Journal of Trading*, 2(4), 82-100.

Vincent, S. (2011) Is portfolio theory harming your portfolio? Green River Asset Management Working Paper, www.ssrn.com/abstract=1840734.

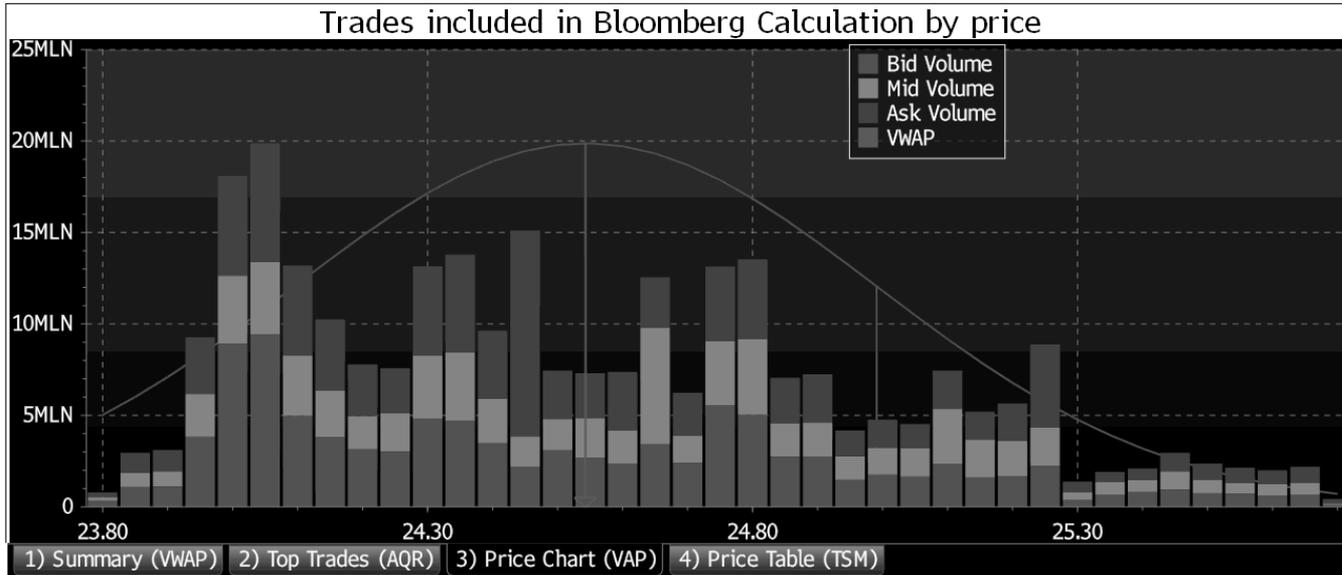

Fig. 1A

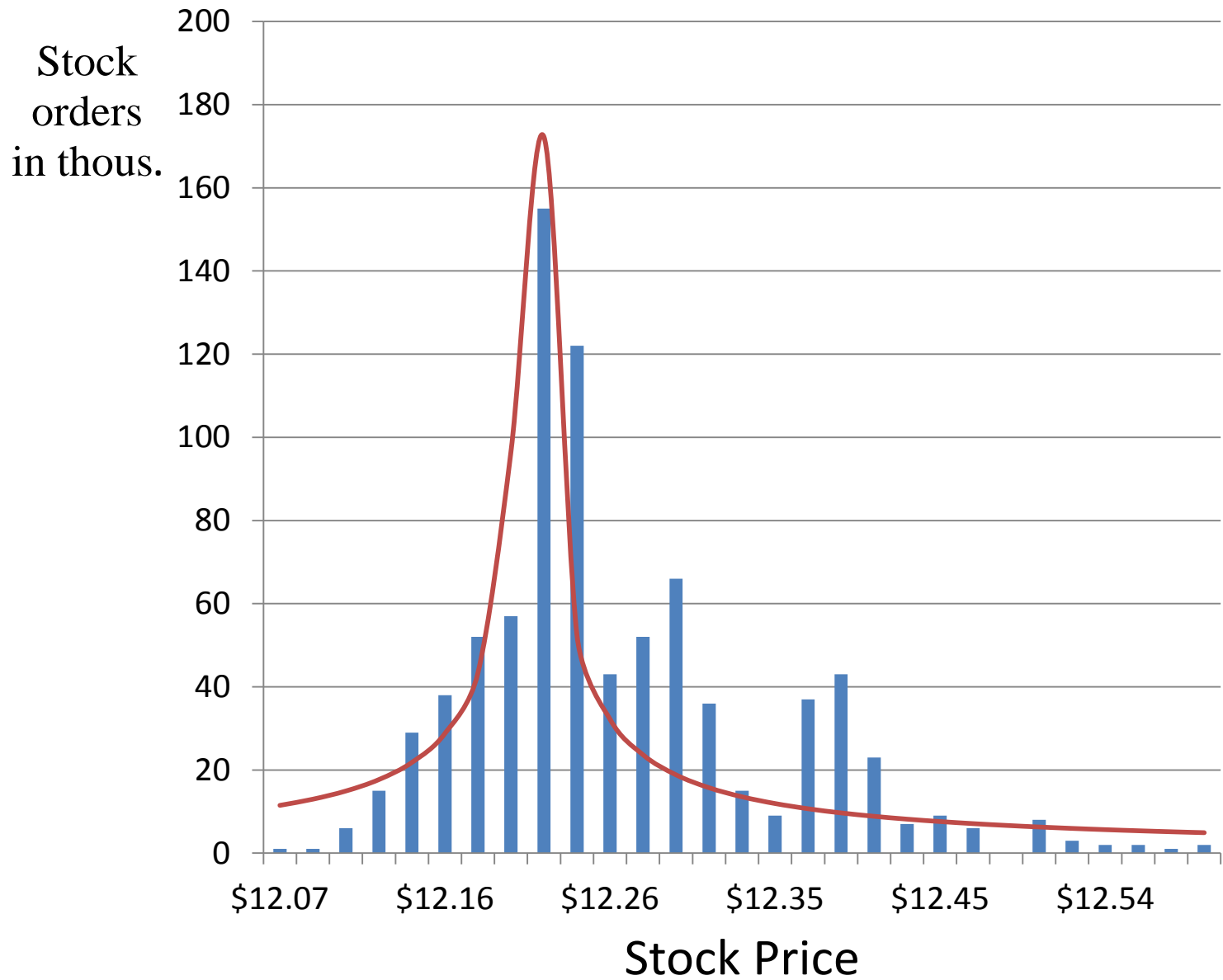

Fig. 1B

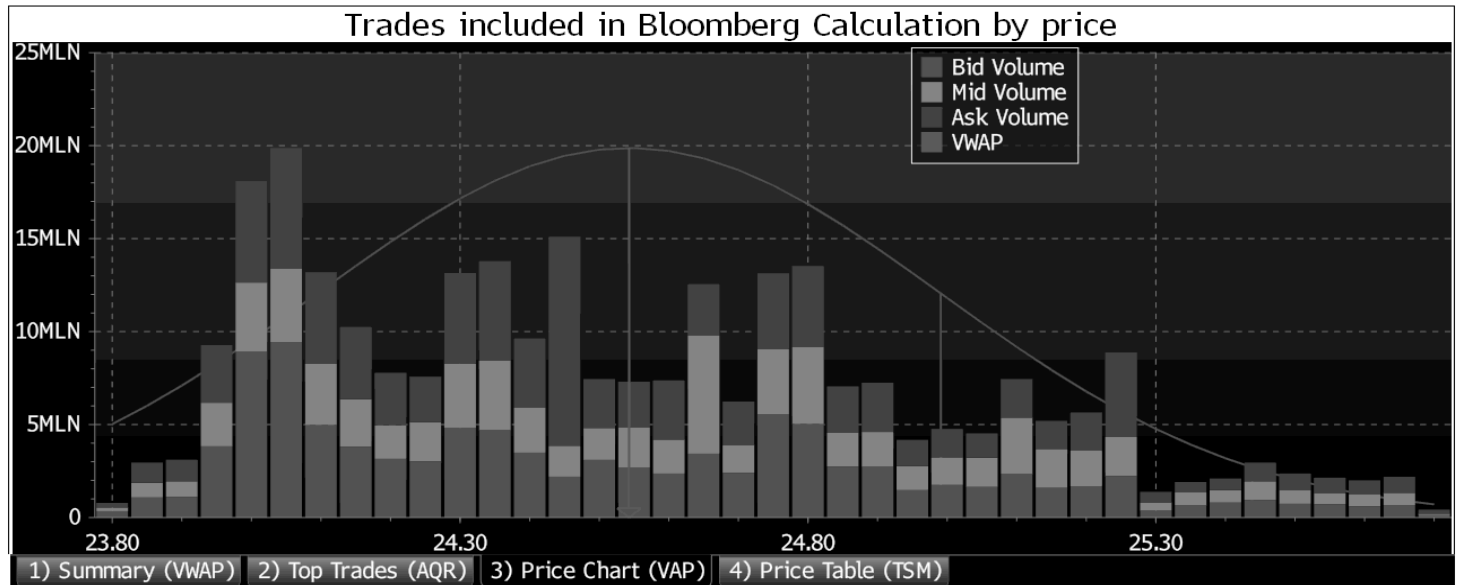

Fig. 2A

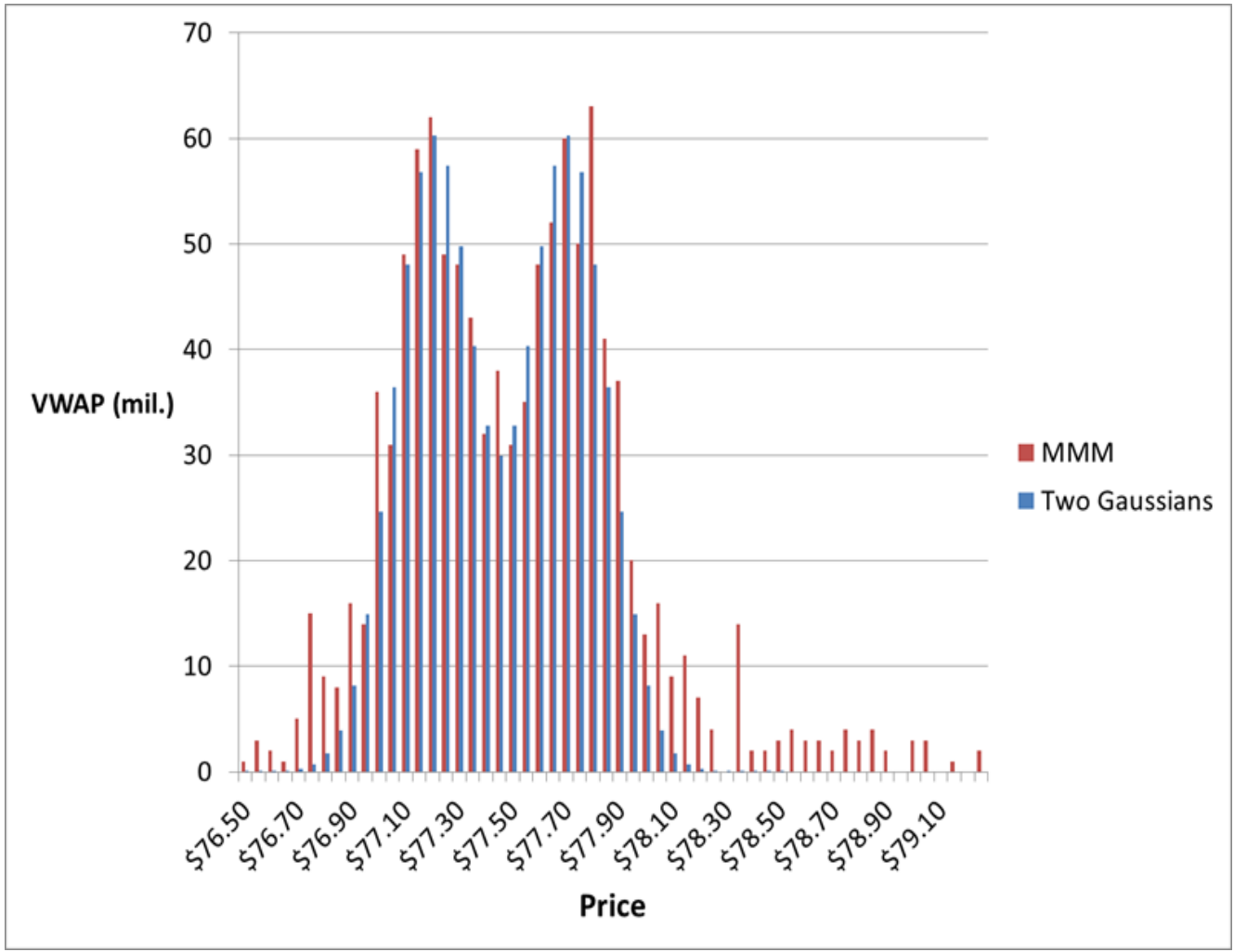

Fig. 2B

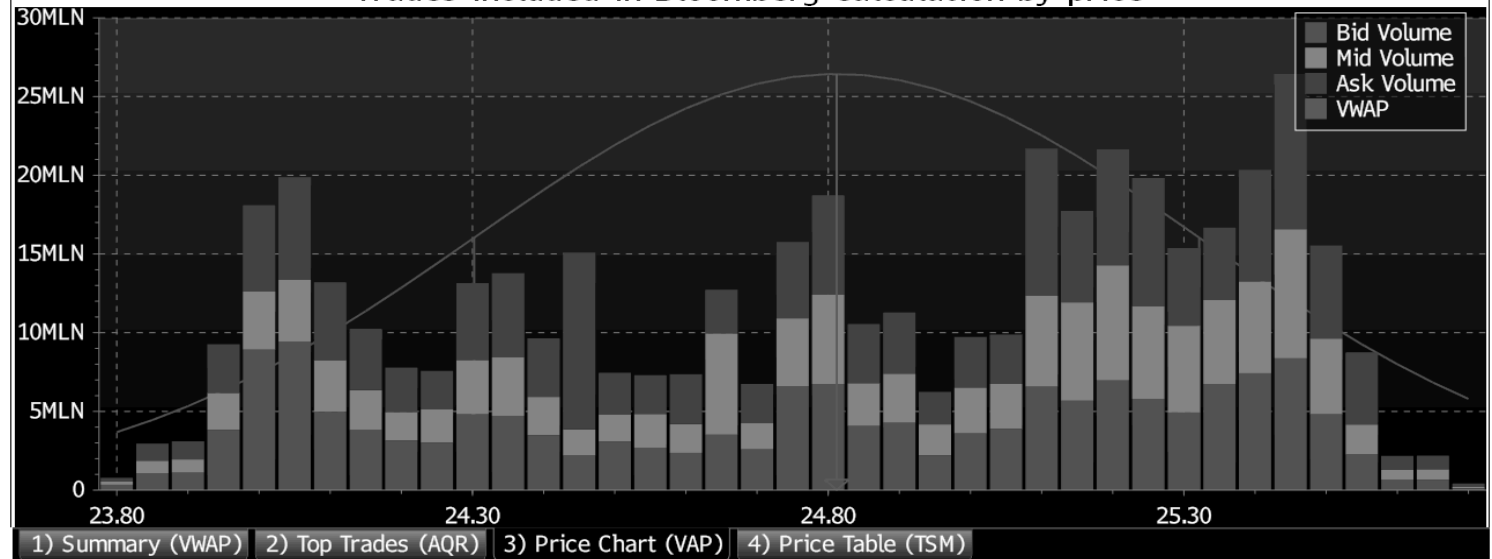

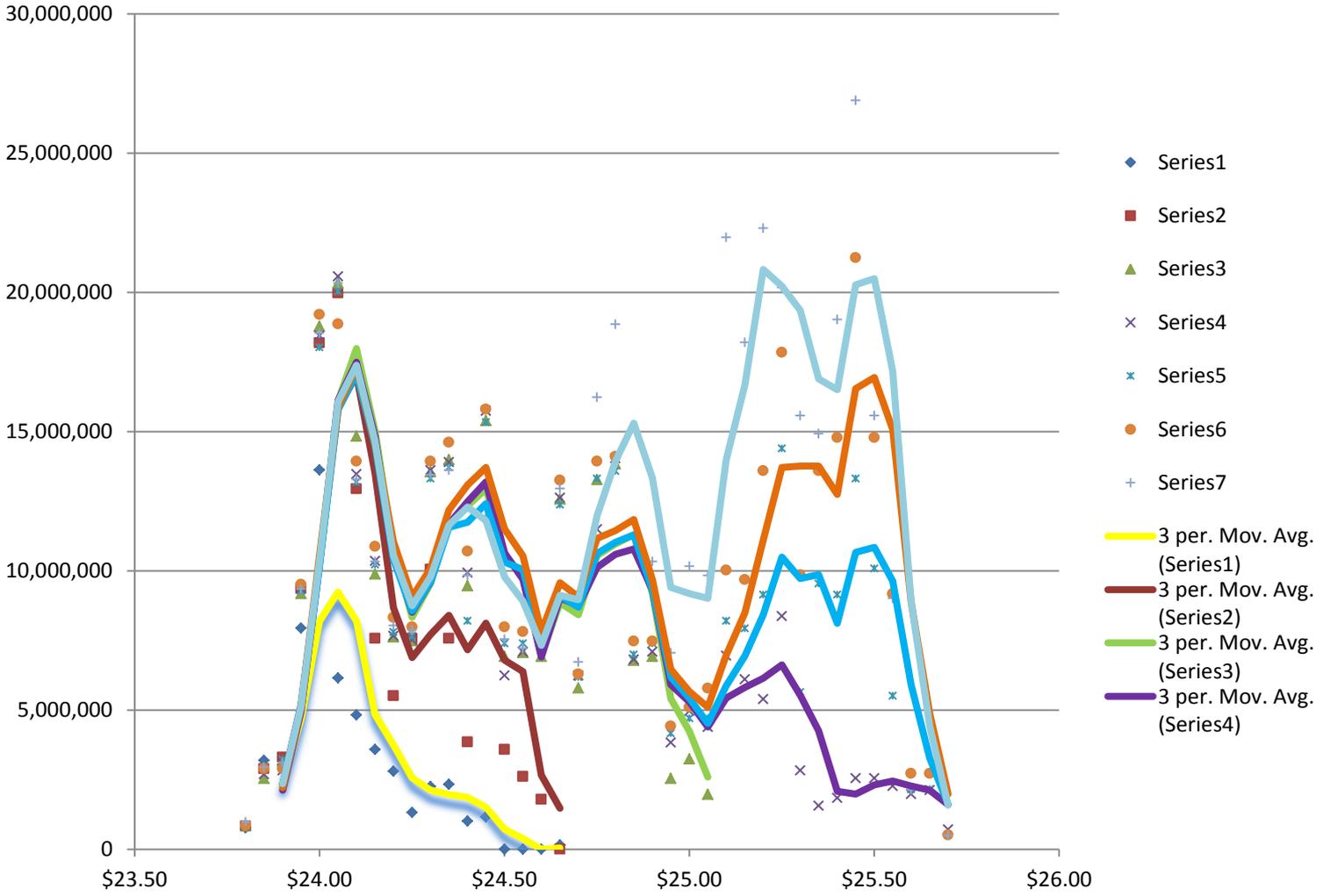

Fig. 3A

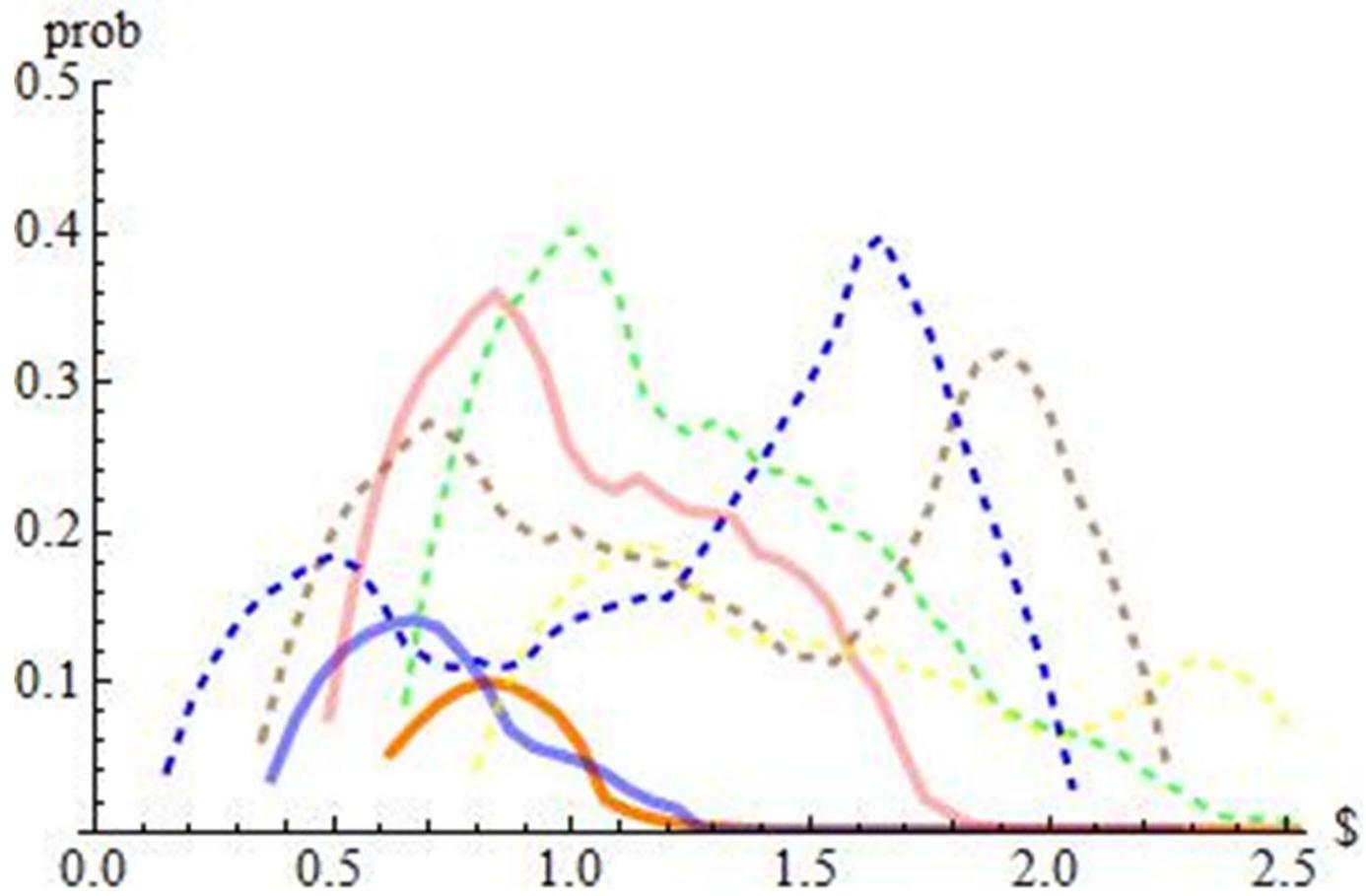

Fig. 4 B

Price deviation at volume

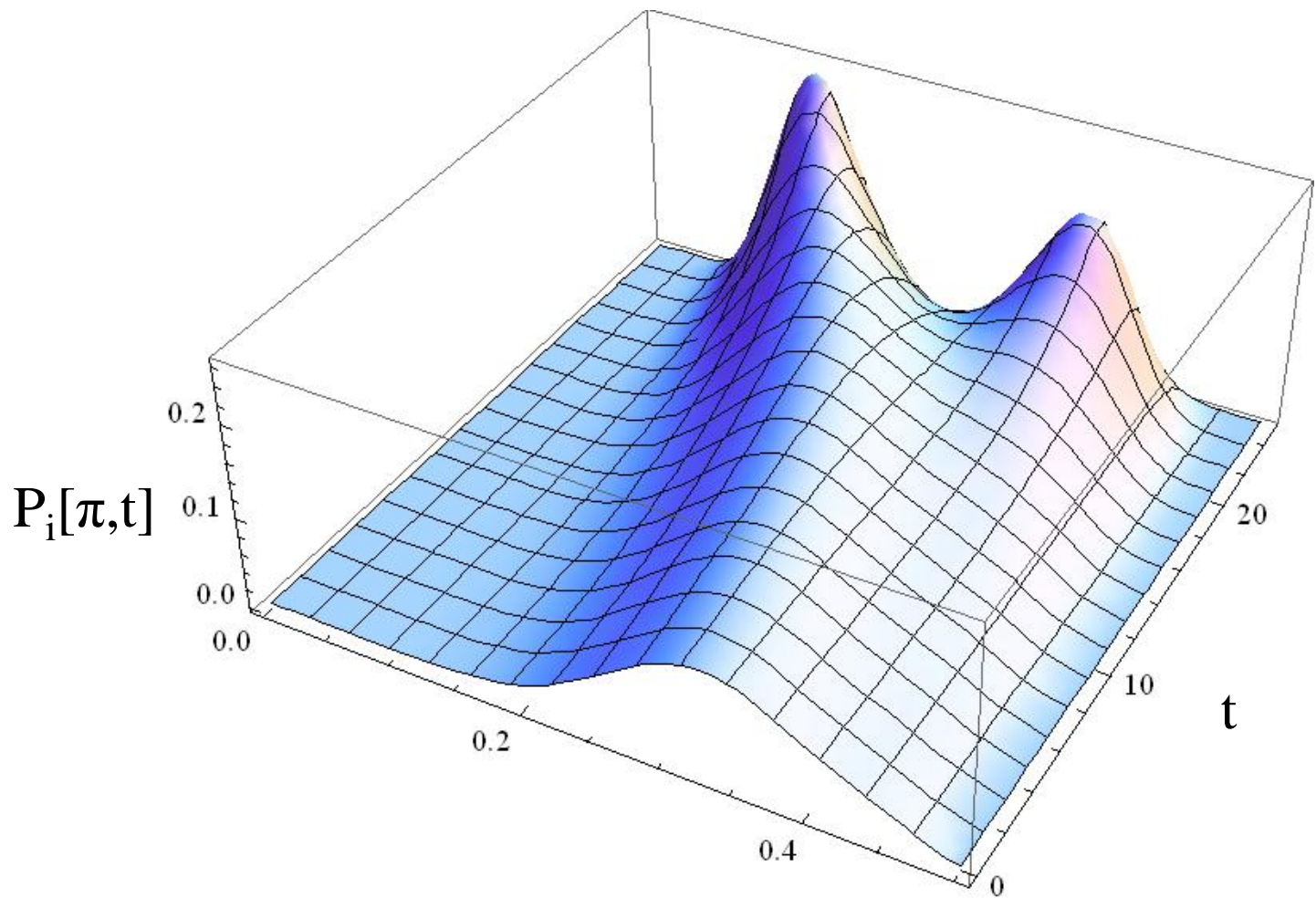

Fig. 4A

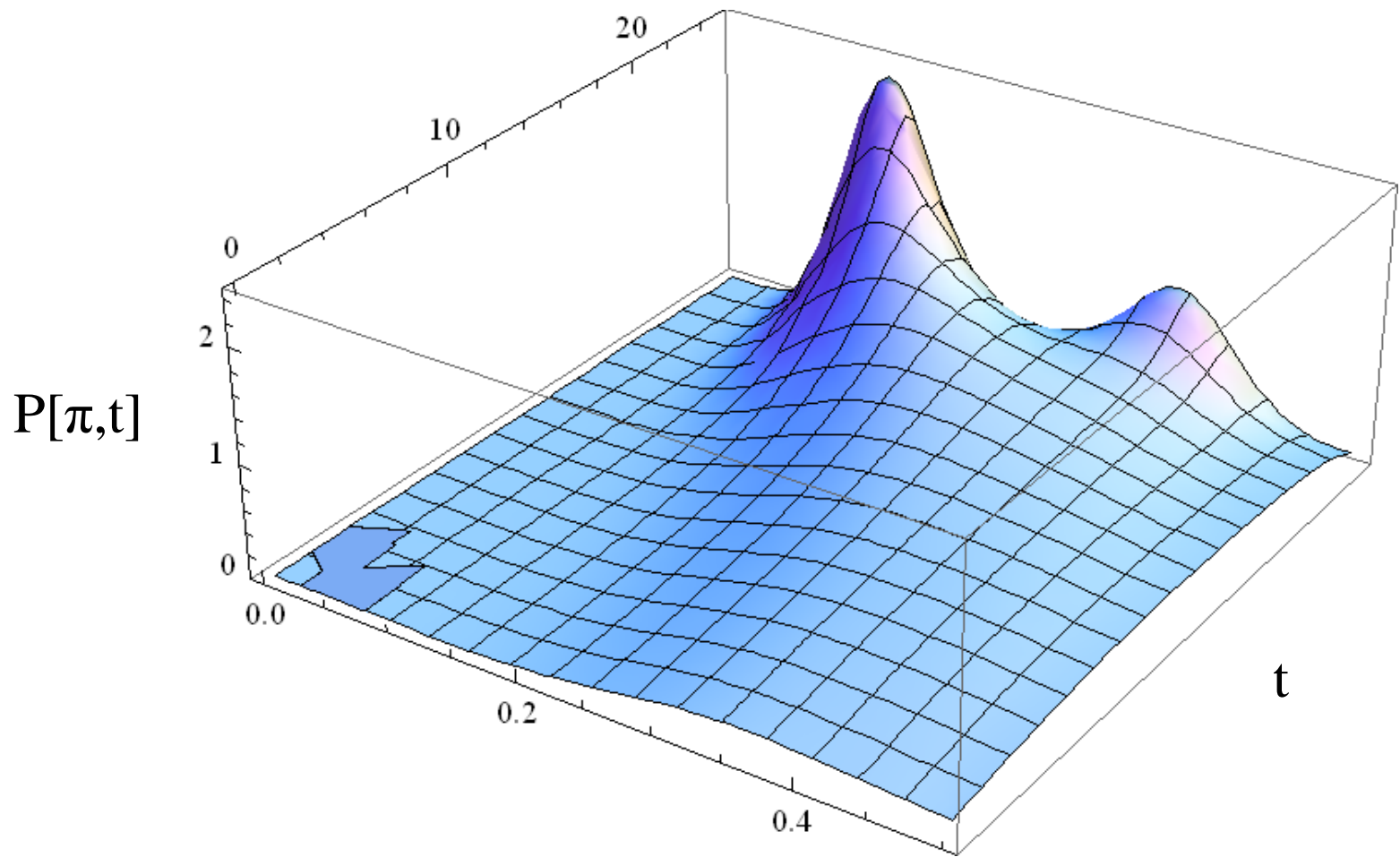

Fig. 4B

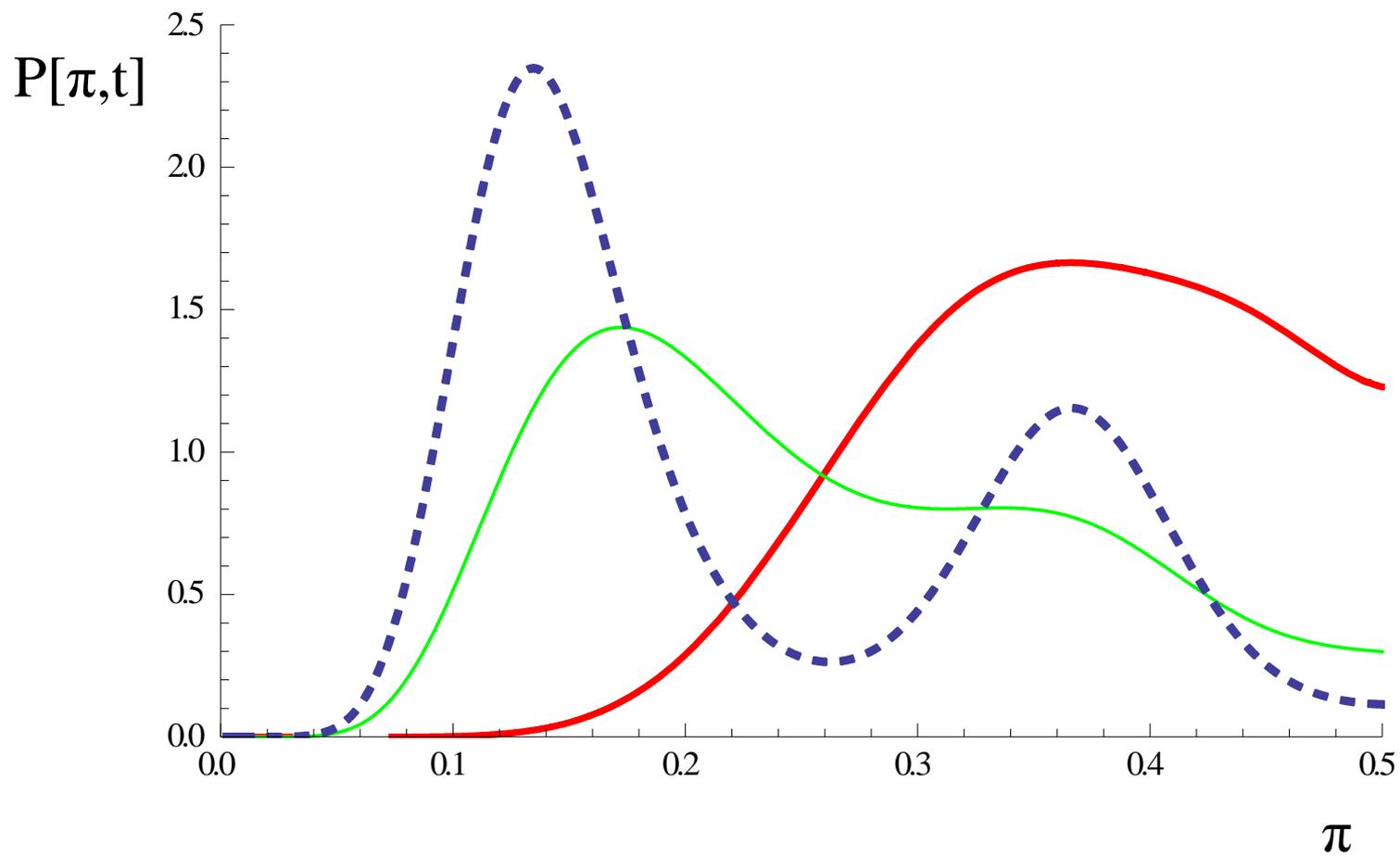